# Atypical magnetic behavior in the incommensurate [CH$_3$NH$_3$][Ni(HCOO)$_3$] hybrid perovskite

Breogán Pato-Doldán,[a,‡] Laura Cañadillas-Delgado,[b,‡] L. Claudia Gómez-Aguirre,[a]

M. A. Señarís-Rodríguez,[a] Manuel Sánchez-Andújar,[a] Óscar Fabelo,[b,c,*] Jorge Mira[d,*]

[a] Department of Chemistry, Faculty of Sciences, Universidade da Coruña, 15071 A Coruña, Spain.

[b] Institut Laue-Langevin, 6 rue Jules Horowitz, BP 156, 38042 Grenoble Cedex 9, France.

[c] Departamento de Física, Universidad de La Laguna, Avenida Astrofísico Francisco Sánchez s/n, 38200 La Laguna, Tenerife, Spain.

[d] Departamento de Física Aplicada and iMATUS, Universidade de Santiago de Compostela, 15782 Santiago de Compostela, Spain.

[‡]These authors contributed equally to this work

*Corresponding author: fabelo@ill.fr
*Corresponding author: jorge.mira@usc.es

*KEYWORDS: Ferromagnetic, hybrid perovskites, formates, incommensurate structure.*

**ABSTRACT:** A plethora of temperature induced phase transitions have been observed in [CH$_3$NH$_3$][M(HCOO)$_3$] compounds, where M is Co(II) or Ni(II). Among them, the nickel compound exhibits a combination of magnetic and nuclear incommensurability below Néel temperature. Despite the fact that the zero-field behavior has been previously addressed, here we study in depth the macroscopic magnetic behavior of this compound to unveil the origin of the atypical magnetic response found in it and in its parent family of formate perovskites. In particular, they show a puzzling magnetization reversal in the curves measured starting from low temperatures, after cooling under zero field. The first atypical phenomena is the impossibility of reaching zero magnetization, even by nullifying the applied external field and even compensating it for the influence earth's magnetic field. Relatively large magnetic fields are needed to switch the magnetization from negative to positive values or vice versa, which is compatible with a soft-ferromagnetic system. The atypical path found in its first magnetization curve and hysteresis loop at low temperatures is the most noticeable feature. The magnetization curve switches from more than 1200 Oe from the first magnetization loop to the subsequent magnetization loops. A feature that cannot be explained using a model based on unbalanced pair of domains. As a result, we decipher this behavior in light of the incommensurate structure of this material. We propose, in particular, that the applied magnetic field induces a magnetic phase transition from a magnetically incommensurate structure to a magnetically commensurate structure.

INTRODUCTION

Coordination polymers are hybrid inorganic/organic structures formed by metal cation centers that are linked by ligands, in the form of one-, two-, or three-dimensional crystalline structures. Such ligands open spaces in the structure with the capacity of hosting diverse cations. This allows a tailoring that has yielded a plethora of potential applications and functionalities[1,2] thanks to their optical,[3] ferroelectric,[4–6] or magnetic properties.[7,8]

An example of these materials are those of formula [amineH][M(HCOO)$_3$] (where M is a divalent transition metal cation) which present an ABX$_3$ perovskite structure, where the metal cations (B = M$^{2+}$) linked by formate groups (X = HCOO$^-$) form the BX$_3$ skeleton, and protonated amine cations (amineH) occupy the cavities (A=[CH$_3$NH$_3$]$^+$, [(CH$_3$)$_2$NH$_2$]$^+$, [CH$_3$CH$_2$NH$_3$]$^+$, [(CH$_2$)$_3$NH$_2$]$^+$, [C(NH$_2$)$_3$]$^+$, [HONH$_3$]$^+$, [NH$_2$NH$_3$]$^+$, etc).[9–12] Some of their magnetic, dielectric and even multiferroic properties have already been described.[13,14] The formate anion linker is a good choice for several reasons: it is a short ligand and it can adopt various bridging modes and extended structures, thus providing significant magnetic coupling between magnetic metal sites.[15]

Typically, the [amineH][M(HCOO)$_3$] compounds display weak ferromagnetic arrangements in the range 8-36 K depending on the specific metal.[16,17] This feature can be explained by the presence of a non-centrosymmetric ligand (formate) between the magnetic ions, which allows the occurrence of

Dzyaloshinsky–Moriya (DM) interactions (antisymmetric exchange), giving rise to spin canting.[18-20]

When a methylammoniun cation is placed inside the cubooctraedral cavities, the resulting [CH$_3$NH$_3$][M(HCOO)$_3$] materials show weak ferromagnetic ordering below $T_N$.[21-24] Within this family, [CH$_3$NH$_3$][Ni(HCOO)$_3$] is especially appealing, as it displays negative susceptibility in some zero-field-cooling magnetization versus temperature curves.[21]

When a magnetic material is under the influence of an external magnetic field, its global magnetic moment tends to align, usually, with the external field. But, in some cases, the alignment of the net magnetization occurs in the opposite direction of the magnetic field (negative magnetization). This behavior has been known since long ago in some ferrimagnets below their compensation temperature.[25] Among them, several families of coordination polymers show negative magnetization, e.g. formates,[26,27] oxalates[28-30] and Prussian blue analogues.[31-36]

In all these cases the effect can be explained by the existence of two different subnets with different net magnetization and ordering temperature. The first subnet forces the second one, due to the antiferromagnetic exchange, to order towards the opposite direction of the applied magnetic field.

Negative magnetization curves have also been observed in weak ferromagnetic systems like LaVO$_3$ or YVO$_3$.[37-39] The mechanism could also be the same in the case of the negative zero field cooling magnetization observed at low fields in Fe$_2$OBO$_3$,[40] but it has also been suggested as an explanation the competition between inter-ribbon versus intra-ribbon exchange interaction of different signs (and temperature dependences), like in the case of Co$_2$VO$_4$ (Co$^{2+}$[Co$^{2+}$V$^{4+}$]O$_4$), arguing that the competition between Co$^{2+}$-O-V$^{4+}$ and direct V$^{4+}$-V$^{4+}$ cants the vanadium and cobalt spins in opposite directions, leading to a compensation point and magnetization reversal.[41]

Nevertheless, in [CH$_3$NH$_3$][Ni(HCOO)$_3$] the mechanism of negative magnetization and the anomalies observed in the hysteresis loops cannot be explained using a non-compensation model between magnetic subnets.

The case of this compound is quite uncommon, and it needs specific circumstances to show up. Cañadillas-Delgado et al. have recently analyzed it by neutron diffraction,[42] and found that this material is a rare case, where structural incommensurability and magnetic incommensurability have both a proper character. In the following sections, its magnetic behavior is deeply analyzed using magnetometry measurements to confirm the proper character of the magnetic incommensurability and to unveil the reasons of its rare magnetic behavior, which could also help to explain similar features in other parent coordination polymers with perovskite structure.

EXPERIMENTAL SECTION
**Materials**

NiCl$_2$ (98%, Aldrich), methylamine hydrochloride (99% Aldrich), sodium formate (≥99%, Aldrich) and N-methylformamide (99%, Aldrich) were commercially available and used as purchased without further purification.

**Synthesis**

A mixture of NiCl$_2$ (1 mmol), NaCHOO (3 mmol), CH$_3$NH$_2$·HCl (1 mmol), methylformamide (8 mL), and H$_2$O (8 mL) was heated in a Teflon-lined autoclave (45 mL) at 140 $^0$C for 3 days. After slow cooling to room temperature, green crystals suitable for single-crystal X-ray crystallographic analysis were obtained. They were collected, washed with ethanol and dried at room temperature. Large single crystals, suitable to carry out magnetic measurements, were obtained by slowly evaporating the mother liquid for about 4 months at room temperature.

**Heat capacity**

Heat capacity as a function of temperature was measured on a single crystal using a Quantum Design PPMS (Physical Properties Measurement System) in the temperature range 1.9 -300 K. The sample was fixed to the sample holder with Apiezon N grease.

**Measurement of magnetic properties**

The magnetic properties were studied in a Quantum Design MPMS SQUID magnetometer in both polycrystalline samples and oriented single crystals. Single crystals were oriented on the X-ray diffraction unit of the University of Santiago de Compostela and mounted on a straw, an error of ± 5 ° along the different orientations could occur. Zero-field-cooled (ZFC) and field-cooled (FC) magnetic susceptibility data were obtained under different magnetic fields in the temperature range 2≤ T (K) ≤ 300. Hysteresis loops in ZFC conditions were obtained at 2 K varying the field up to ± 50 kOe. The data were corrected using Pascal's constants to calculate the diamagnetic susceptibility, and an experimental correction for the sample holder was applied. In order to reduce the remnant field in the magnet of the magnetometer, the "reset field" option of the system was used before lowering the temperature for a ZFC curve. Also, in order to test the influence of the remnant field of the magnet (even after the resetting of the field) the ultralow field option has been used in some specific measurements, which nulls the applied magnetic field at the specific position of the sample in the chamber. For this, the residual field was measured with a fluxgate magnetometer and then a compensating field was applied in the superconducting solenoid to null the residual one. Even after this operation, a field closer to zero than -0.1 Oe could not be assured in all the relevant parts of the chamber.

RESULTS
**Crystal and magnetic structure**

Although the description of the crystal and magnetic structure of [CH$_3$NH$_3$][Ni(COOH)$_3$] is not the main objective of this work, to better understand the magnetic behavior of this compound a brief description of the nuclear and magnetic structure based on the neutron single-crystal diffraction data will be discussed in this section.

[CH$_3$NH$_3$][Ni(COOH)$_3$] shows a structural phase transition at 84 K (see Fig. 1 for an overview of the phases of this system), involving a transformation from the commensurate *Pnma* space group to the incommensurate *Pnma* (00$\gamma$)0s0 space group with wave vector **q** = 0.1426(2)*c**. The average structure, described in the *Pnma* space group, is distorted by the application of a modulation function that exhibits a sinusoidal behavior. The amplitudes of the displacive modulation are mainly applied over the *b*-axis with a small component in the *a*- and *c*-axis for some atoms.

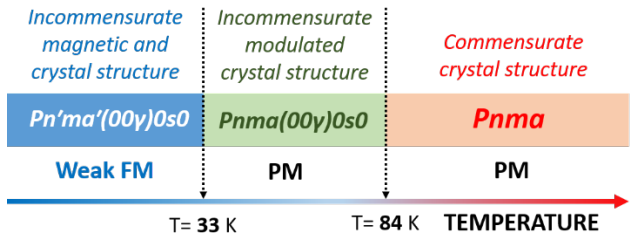

Figure 1. Summary of the phase transitions with temperature in [CH$_3$NH$_3$][Ni(HCOO)$_3$]: from paramagnetism (PM) to weak ferromagnetism at 33 K and from incommensurate to conmensurate crystal structure at 84K.

The application of these modulations induces distortions on the framework and counter-ions. However, the local structure is not strongly affected. In the case of the NiO$_6$ octahedron, these distortions produce a displacement of the Ni atom along the *b*-axis of ca. 0.3 Å; however, the local environment remains octahedral with values of Ni-O distances between 2.054(2) and 2.076(2) Å.

There is a significant change in the hydrogen-bonded network between the commensurate and the incommensurate phases. The hydrogen atoms of the CH$_3$ group do not establish any hydrogen bond, neither in the commensurate nor in the incommensurate phase. However, two of the three hydrogen atoms of the NH$_3$ group clearly set hydrogen bonds with the nearest oxygen atoms of two formate groups, in both phases. But in the incommensurate phase the third hydrogen atom of the NH$_3$ group fluctuates following a sinusoidal distortion between two oxygen atoms from the same formate ligand: this interaction is the most probable origin of the structural modulation.

The combination of displacive and magnetic modes is needed to fit the experimental data. The determination of the compatible superspace magnetic groups has been done combining two independent modulation vectors, the **q** = 0.1426*c**, which accounts for the structural distortions, and **k** = (0, 0, 0), over the previously distorted structure, and therefore, incommensurate from the structural point of view.

The symmetry analysis reveals that below 33 K [CH$_3$NH$_3$][Ni(COOH)$_3$] can be described using the *Pn'ma'*(00γ)0*s*0 super-space group, with origin (0,0,0).

This super-space group allows twelve free modes for the magnetic atoms, which are divided on strain, displacive and magnetic modes. The strain models are considered during the indexing procedure and, therefore, for the magnetic analysis they can be discarded. The second group represents three displacive modes, which are responsible for structural modulation (atomic displacement). The last group involves six pure magnetic modes, three for the *x*, *y* and *z* components of the homogeneous moment, and three sinusoidal modulations with amplitudes along *x*, *y* and *z*, being these last three modes the responsible of the proper magnetic contribution.

Below 33 K, the refined model can be described as chains oriented ferromagnetically along the *c* axis, and antiferromagnetically coupled with the adjacent along the *a* and *b* directions (C-type antiferromagnet). The magnetic moments are tilted along the *b* axis, giving rise to a non-zero ferromagnetic component along this direction (Fig. 2). This contribution arises from the non-zero value of the *y* component of the homogeneous moment, and therefore this component has an improper origin.

Furthermore, the development of proper incommensurability modes promotes the modulation of the orientation of the magnetic moments, exhibiting a sinusoidal behavior with the main contribution to the magnetic modulation amplitude along the *a* axis. The amplitude of the magnetic modulation along the *b* axis is zero within the experimental error, and the amplitude along the *c*-axis is almost four times weaker than along the *a* axis. The effect of these two modulation components produces the non-displacive modulation of the magnetic moment in the *ac* plane (Fig.2). The contribution of these two components is important for explaining the atypical hysteresis loops, as it will be shown subsequently in the text. The obtained value for the Ni(II) magnetic moment is in average 2.15(7) μ$_B$, a small modulation in the magnetic moment modulus is observed, however the variation (2.14 to 2.16 μ$_B$) is within the error bar of our refinement.

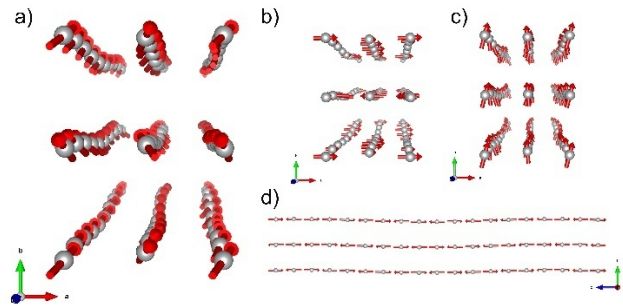

Figure 2. (a) Magnetic structure at 5K and zero-field of [CH$_3$NH$_3$][Ni(COOH)$_3$]. (b) magnetic moment component along the *a* axis (red), showing that along this direction the system is AF. (c) magnetic moment component along the *b* axis (green), where there is a global weak ferromagnetic component. (d) Component of the magnetic moments along the *c* axis (blue), that corresponds with the main component and presents a global AF behavior. The magnetic moment amplitudes for the components along the *a* and *b* axis have been multiplied by two for the sake of clarity. The structure represented in a) is the sum of b), c) and d).

**Magnetic properties**

Fig.3 shows the ZFC-FC susceptibility χ(T) of a polycrystalline sample of [CH$_3$NH$_3$][Ni(HCOO)$_3$] under an applied field of 1000 Oe, where an increase of the magnetization and divergence of the ZFC-FC curves is visible below T ≈ 33 K. The most interesting feature is, however, a ZFC curve that shows negative susceptibility until 25 K.

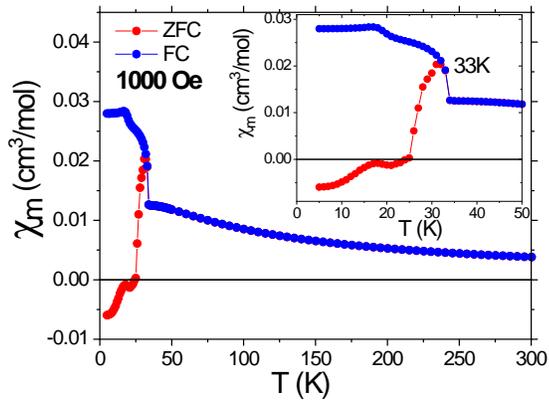

Figure 3. ZFC-FC susceptibility χ(T) of a polycrystalline sample of [CH$_3$NH$_3$][Ni(HCOO)$_3$] under an applied field of 1000 Oe.

The linear fit of the susceptibility data above 50 K provides a good agreement to a Curie–Weiss behavior from which a value of θ = −65 K and a μ$_{eff}$ = 3.35 μ$_B$ are calculated (see Table 1).

The latter is close to the expected for a Ni$^{2+}$ cation (d$^8$) with S=1 and g=2.00 (μ$_{teo}$ =2.83 μ$_B$). The negative value of θ implies that the main exchange interaction is antiferromagnetic. The magnetic transition at 33 K is also detected in a peak in the thermal dependence of the specific heat at that temperature (Fig. 4).

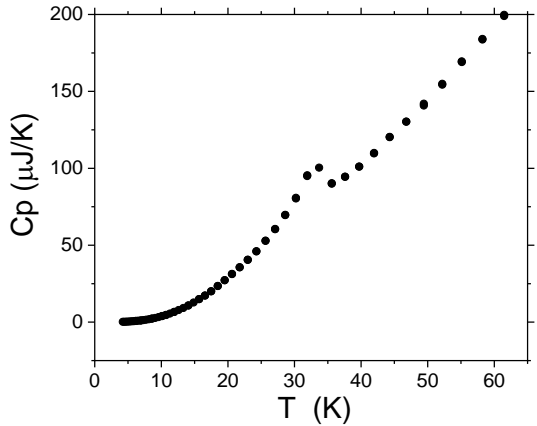

Figure 4. Specific heat capacity (Cp) as a function of the temperature for [CH$_3$NH$_3$][Ni(HCOO)$_3$], where the peak associated to the magnetic transition point is visible.

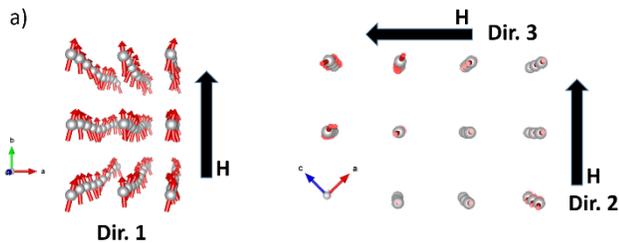

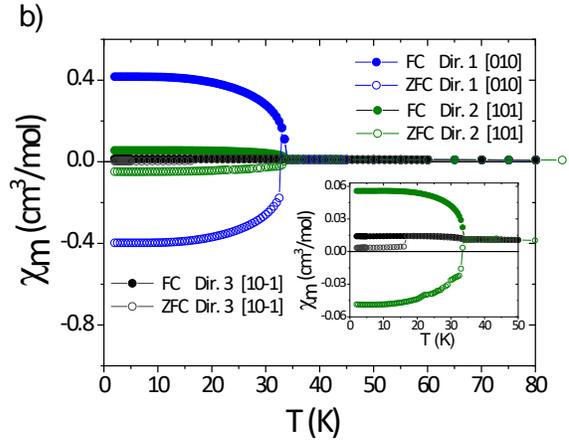

Figure 5. (a) Orientations along where the magnetization has been measured for [CH$_3$NH$_3$][Ni(HCOO)$_3$]. (b) Thermal dependence of χ$_m$ for [CH$_3$NH$_3$][Ni(HCOO)3] along the orientations [010], [101] y [10-1] and under a field of 100 Oe.

When the ZFC-FC susceptibility χ(T) is measured in single crystals (along the directions indicated in Fig. 5a) for a lower magnetic field (100 Oe), an almost specular behavior between ZFC and FC curves is clearly observed below T ≈ 33 K, with negative magnetization values more evident than before for the ZFC curve (Fig. 5b).

In order to discard any artifact or a possible extrinsic origin of this phenomenon, we examined more in detail our measuring procedures. First, we performed FC magnetization curves under small magnetic fields, i.e., magnetization vs. temperature curves by cooling the samples under low magnetic fields. We observed then that, when applying a cooling field of -5 or -10 Oe, the ZFC branch of the susceptibility curve measured at 100 Oe (shown in Fig. 6) was similar to that in Fig.5b, but, when this cooling field was +5 or +10 Oe, the measured curve at 100 Oe showed an almost specular behavior: it completely reversed its sign with respect to the previous case.

This means that the system is extremely sensitive to any trapped magnetic field in the magnet: depending on the sign of this trapped field that acts on the sample while cooling, the magnetization defines its sign once the temperature goes below 33 K. This is a crucial point when facing the study of the magnetism of this system; we join the opinion of Belik[43,44] and Kumar and Sundaresan,[45] who have warned about this tricky circumstances in the measurement of the magnetizations of CoCr$_2$O$_4$ or classical materials in solid state physics like the perovskites BiFeO$_3$-BiMnO$_3$ and YVO$_3$.

Actually, even after using the ultralow field option of our magnetometer to reduce the trapped field acting on the sample chamber, we could not get rid of this problem: fields as low as ± 0.1 Oe (which is around 1/5 of Earth's magnetic field) were enough to completely polarize the magnetization in one direction or the other. Therefore, we conclude that the negative magnetization found after conventional ZFC procedures arises from the small trapped fields in the superconducting magnets during cooling, that changes drastically the behavior of the material in the ordered phase. It is to be explained, nevertheless, why this happens. We suspect that a similar reason could underlie many of the negative magnetizations reported in similar materials,[46–48] as trapped fields in conventional SQUID magnetometers are usually negative.

It is worth mentioning that the application of moderate magnetic fields (i.e., 100 Oe) during the heating of the sample is not strong enough to reorient the magnetization on the sample in the direction of the magnetic field at some point of the heating curve. For 1000 Oe, the ZFC of the polycrystalline sample switches from negative to positive at around 25 K (Fig. 3).

Hysteresis loops performed along the different directions are quite illustrative about the magnetic behavior of this system (Fig. 7). Especially appealing is the result along the [010] direction and, moreover, the exotic initial magnetization curve (Fig. 7c, green line).

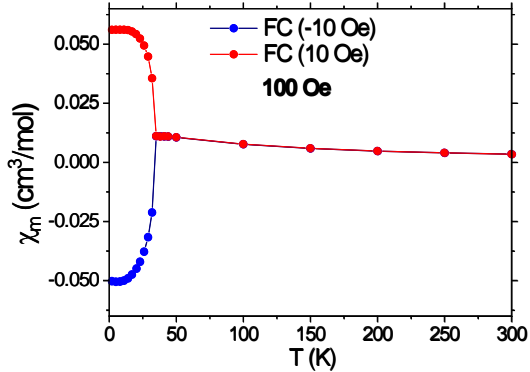

Figure 6. Thermal dependence of $\chi_m$ for $[CH_3NH_3][Ni(HCOO)_3]$ for a polycrystalline sample under a field of 100 Oe, after being field-cooled (FC) under +10 Oe and -10 Oe.

First of all, it seems impossible to set to zero the magnetization of $[CH_3NH_3][Ni(HCOO)_3]$ below 33 K along the [010] direction. In Fig. 7c it is seen that the initial magnetization curve starts from a negative value (due to a trapped field between -0.1 and 0 Oe), it switches to positive at 500 Oe and, later on, the magnetization switches its sign at ~ +/- 1.7 T, creating a hysteresis loop. But the absolute value of M from which it jumps at 500 Oe to positive at the first run is the same as at the following runs, i.e. it seems that this value corresponds to the fully magnetized sample. Why then at the first run it jumps from this state to positive at 500Oe, but at the next rounds it jumps from practically the same state but at 1700 Oe and not at 500Oe? If the first jump from negative to positive at 500 Oe occurred from a value of M smaller that the value at next runs, one could understand why it also occurs at a smaller field, at 500 Oe. The question is why it jumps from the same value, but at different fields.

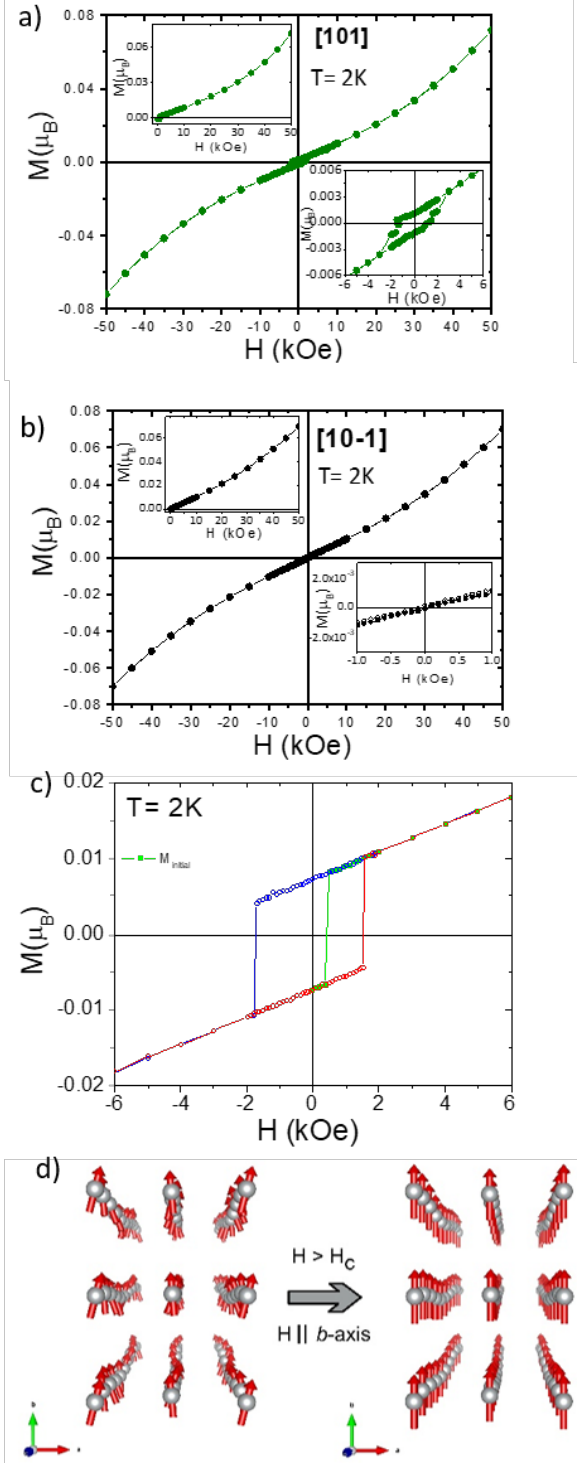

Figure 7. Isothermal magnetization of $[CH_3NH_3][Ni(HCOO)_3]$ along a) [101], b) [10-1] and c) [010] at 2 K. In c), the initial magnetization at 0 Oe starts in negative values (green line), then it jumps to positive values at around 500 Oe and, from that moment on, it follows a symmetrical hysteresis loop. d) Effect of the suppression of proper magnetic modulations in the ferromagnetic component of the magnetic structure for fields larger than Hc (500 Oe at 2K).

**Discussion**

The behavior of the first magnetization observed in Fig. 7 confirms that the magnetic ground state determined through neutron diffraction experiments slightly changes when the applied magnetic field is above a critical value (500 Oe at 2K). Therefore, an explanation to this effect needs to be found. In case the system was a classical spin canting system, as initially described, then the jump would always occur at the same applied magnetic field, as the energy to move the ferromagnetic component of the canting from negative to positive and vice versa is always the same. The fact that the first magnetization of the hysteresis loop follows a different path in subsequent magnetizations, can be ascribed to the change into the magnetic ground state at higher fields.

Another interesting feature of this system is that it is not possible to set the magnetization to zero below 33 K. This can be explained by the influence that the magnetic field has over the sample: small magnetic fields align the ferromagnetic component along this applied field, decreasing the contribution of the magnetic domains. In a classic system with spin canting, the first magnetization goes to zero because when the temperature is lowered magnetic domains appear. These domains are randomly distributed and as a result they cancel each other and the net magnetization is zero. This does not happen again after the first magnetization, because the applied field forces the ferromagnetic moments to follow the direction of the applied field.

Another key in the analysis of $[CH_3NH_3][Ni(HCOO)_3]$ is that the negative and positive branches of the magnetization curves are not completely specular (Figs. 3, 5 and 6). From this, it can be concluded that the system cooled down at zero field has also magnetic domains. However, these domains are not completely compensated as it occurs in classical systems, In this case, the ferromagnetic component can be modified with very weak magnetic fields, as can be seen from Figure 6, this feature precludes a complete domain compensation.

One must keep in mind that the ferromagnetic component of the incommensurate magnetic structure is not completely aligned along the $b$ axis. There are small perpendicular components (based on the refined amplitudes of the proper magnetic modes, the main component is along the $a$ axis), that provokes the appearance of an incommensurability that can be seen as a "pendulum" of the magnetic moment on the $ab$ plane (see Figs. 2c or 7d). At zero-field, the population of magnetic domains is unbalanced resulting in a non-negligible difference in magnetic susceptibility between the FC and ZFC measurements. However, when a small field is applied (<< 500 Oe), the magnetic domains tend to align with the external magnetic field, giving as a result a single domain system, where the susceptibility curve is symmetrical (Fig. 5).

Therefore, as described above, the unbalance of the magnetic domains cannot be at the origin of the observed jump in the first magnetization at $H_c$ (ca. 500 Oe). This jump must imply a change in the ground state of the magnetic structure. Since the effect is mainly observed when the field is applied in the $b$ direction, this suggests that the ferromagnetic component observed at zero field remains in that direction in the under-field measurements for fields above 500 Oe. Once the critical field is overcome, the hysteresis loop presents a coercive field ca. 1.7 T, which is more than 3 times the critical field of the first magnetization. This coercive field is unbiased neither to positive nor to negative fields, which suggests that the zero-field-to-infield transition is not reversible. This irreversibility together with the relatively large coercive field suggests that the ferromagnetic component of the magnetic moments is better aligned with respect to the $b$ axis. In order to achieve this configuration, a transition from a magnetically incommensurate structure to a magnetically commensurate structure is expected. In other words, the proper magnetic modulations are suppressed leaving only active the improper modes, which are linked to the structure's displacements (see Fig. 7 d). After the first magnetization, this structure remains invariant with only the expected flip of sense of the ferromagnetic component when the applied magnetic moment changes from positive to negative overcoming the coercive field and vice versa.

**CONCLUSIONS**

We have observed that the $[CH_3NH_3][Ni(HCOO)_3]$ hybrid perovskite presents a series of uncommon features. The nuclear transition from commensurate to incommensurate phases induced by the temperature is triggered by the changes in the hydrogen bond network. At lower temperatures, below Néel temperature, it presents a both proper magnetic and nuclear incommensurability; this intricate magnetic structure occurs as a consequence of competing interactions. From the macroscopic magnetic measurements, we have observed that this compound shows several uncommon magnetic responses with non-zero susceptibility after ZFC procedures. This response is due to the non-compensation of the magnetic domains. The unbalance is caused by the small trapped fields in the superconducting magnets during the cooling of the sample. The magnetic field required to orient the ferromagnetic domains in the direction of the applied magnetic field is so weak that the competition of the trapped fields in the superconducting magnets and the magnetic field created by the sample itself precludes the compensation of magnetic domains.

When the hysteresis loops of this compound have been studied, surprisingly, the first magnetization follows a different pathway than subsequent magnetizations. However, although there is a change of sign in the magnetisation, the curve continues smoothly, without showing a drastic change in the magnetisation values in subsequent magnetisations. This effect is most visible when the field is applied in the $b$ direction, the direction in which the ferromagnetic component is pointing. That involve that the ferromagnetic component observed at zero field remains mainly in the same direction after the switch above the critical field (500 Oe). Therefore, a drastic spin reorientation as in a spin-flop system is discarded. After this first magnetization the hysteresis loop exhibits a coercive field, which is more than three times the critical field of the initial magnetization. This coercive field is unbiased neither to positive nor to negative fields, which suggests that the zero-field-to-infield transition is not reversible. Furthermore, the reproducibility in the subsequent cycles rules out that it is an experimental effect (i.e. sample movement, etc.).

Considering the nuclear and magnetic structure obtained at zero field, the most plausible explanation is that the influence of the applied external field suppressed some of the active magnetic modes at zero field. The suppression of the proper magnetic modes slightly modifies the magnetisation values, which explains why the magnetisation cycles have no other feature than a change in sign at the critical field, but the resulting mag-netic structure, a magnetically commensurate structure, is energetically more favourable, which explains the increase in the value of the coercive field.

### Funding Sources

The authors thank financial support from Ministerio de Economía y Competitividad MINECO and EU-FEDER (projects MAT2017-86453-R and PDC 2021-121076-I00).

### ACKNOWLEDGMENT

We are grateful to Dra. Ana Arauzo at Servicio de Medidas Físicas of the Universidad de Zaragoza for heat capacity data. O.F. acknowledges the Spanish Ministry of Universities (UNI/551/2021) and the European Union through the Funds Next Generation.

### ABBREVIATIONS

ZFC Zero Field Cooling, FC Field Cooling, SQUID Super Quantum Interferometer Device.

12050-12064.

Table 1. Summary of the results of the fitting to the Curie-Weiss and Lines model for a single cristal of [CH$_3$NH$_3$][Ni(HCOO)$_3$] along the directions [010], [101] y [10-1], compared against a polycrystalline sample measured at 1000 Oe.

|  | Polycrystalline sample | Single cristal | | |
| --- | --- | --- | --- | --- |
|  |  | [010] | [101] | [10-1] |
| T$_{C-W}$[a] / K | 50-300 | 68-300 | 80-300 | 70-300 |
| R$^2$ (CW) [b] | 0,9998 | 0,9996 | 0,9998 | 0,9995 |
| C / cm$^3$Kmol$^{-1}$ | 1,40 | 1,13 | 1,17 | 1,10 |
| θ / ° | -64,96 | -51,6 | -53,7 | -48,6 |
| μ$_{eff}$ / μB | 3,35 | 3,01 | 3,07 | 2,97 |
| T$_L$[c] / K | 65-300 | 80-300 | 62-300 | 70-300 |
| R$^2$ (L) [d] | 0,9997 | 0,998 | 0,998 | 0,997 |
| J / cm$^{-1}$ | -9,38 | -7,45 | -8,60 | -7,86 |
| g | 2,33 | 2,11 | 2,16 | 2,09 |

[a] Temperature interval for the fitting to the de Curie-Weiss model.
[b] Goodness of fit of the Curie-Weiss fitting
[c] Temperature interval for the fitting to the de Lines model
[d] Goodness of fit of the Lines fitting